\documentclass[aps,prl,twocolumn,showpacs,superscriptaddress,groupedaddress]{revtex4}
\usepackage{graphicx}
\usepackage{bm}        
\usepackage{amssymb}   
\begin{document}

\title{Obstruction enhances the diffusivity of self-propelled rod-like particles  }

\author{Hamidreza Khalilian}
\affiliation{Department of Physics, Institute for Advanced
Studies in Basic Sciences (IASBS), Zanjan 45137-66731, Iran}

\author{Hossein Fazli}
\email{fazli@iasbs.ac.ir} \affiliation{Department of Physics,
Institute for Advanced Studies in Basic Sciences (IASBS), Zanjan
45137-66731, Iran} \affiliation{Department of Biological Sciences,
Institute for Advanced Studies in Basic Sciences (IASBS), Zanjan
45137-66731, Iran}

\date{\today}

\begin{abstract}
Diffusion of self-propelled particles in the presence of randomly distributed obstacles in three dimensions is studied using molecular dynamics simulations. It is found that depending on the magnitude of the propelling force and the particle aspect ratio, the diffusion coefficient can be a monotonically decreasing or a non-monotonic concave function of the obstructed volume fraction. Counterintuitive enhancement of the particle diffusivity with increasing the obstacles crowd is shown to be a combinatory effect of the self-propelling force and the anisotropy in the shape of the particle. Regions corresponding to monotonic and non-monotonic dependence of the particle diffusivity on the obstacle density in propelling force-aspect ratio plane are specified theoretically and using the simulation results.
\end{abstract}

\maketitle


Objects under simultaneous influence of a self-propelling
\begin{figure*}
\includegraphics[width=1.8 \columnwidth]{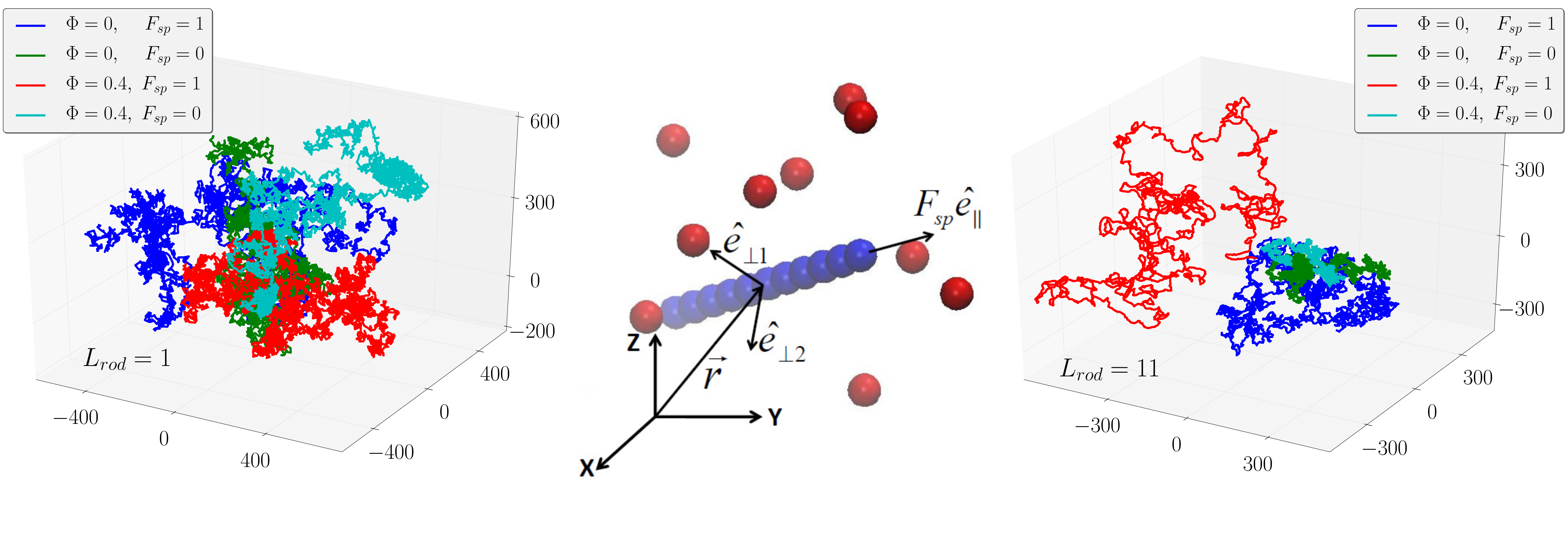}
\caption{(Color online) Schematic of a rod-shaped particle and spherical obstacles in three dimensions. Shish-kebab model of the rod is used for calculation of its excluded volume interaction with the obstacles (middle). Sample trajectories of spherical ($L_{rod}=1$) and rod-shaped ($L_{rod}=11$) particles (left and right). The most diffusivity of the rod corresponds to its obstructed SP diffusion. } \label{fig1}
\end{figure*}
force and a randomly varying force exerted by environment have been of great interest in recent years.
Microscopic organisms \cite{berg1983, okubo1980}, artificial microswimmers \cite{Sowimmer1, Sowimmer2, Sowimmer3, Sowimmer4, Sowimmer5, Sowimmer6, Sowimmer7, Sowimmer8}, synthetic motile objects from molecular scale to microns \cite{AngewChem}, surface-active colloidal particles \cite{s-active1, s-active2} and millimeter-sized manmade particles \cite{Manmade1, Manmade2, Manmade3, Manmade4, Manmade5, Manmade6, Manmade7, Manmade8, Manmade9} are examples of such objects.
Combination of the two mentioned forces has been shown to cause a variety of regimes in dynamics of such objects \cite{s-active2,Peruani910}.

The shape of a particle dispersed in a fluid strongly affects its diffusion dynamics. An anisotropic particle experiences different values of friction coefficient when it moves in different directions and the coupling of translational and rotational motions makes understanding and visualization of its diffusion noticeably difficult \cite{Han}. Self-propulsion amplifies the difference between dynamics of isotropic and anisotropic particles.



Obstructed diffusion -- diffusion of particles in the presence
of fixed obstacles -- is a ubiquitous phenomenon in nature.
The diffusing object in the presence of obstacles could be a simple spherical particle, a rigid anisotropic
particle or a long flexible polymer. Obstructed diffusion of passive particles of all the mentioned types has been
studied extensively \cite{Mackie, Mercier, Saxton2, Wedemeier, Amsden1, Sung, Muthukumar1, Yamakov2, Chang, Sakha}.
Presence of fixed obstacles, also, has different effects on the diffusion dynamics of isotropic and anisotropic particles. In the latter case, the obstruction effect depends on the direction of motion; a rod-like particle experiences the least (the most) number of collisions with the obstacles when it moves along (perpendicular to) its axis.

For self-propelled (SP) particles of anisotropic shape (such as rod), an important question is how the crowd of fixed obstacles affects their dynamics. In this paper, we study dynamics of rod-shaped SP particles in the presence of randomly distributed spherical obstacles using the results of molecular dynamics (MD) simulations.
We find that depending on the magnitude of the propelling force and the particle aspect ratio, the diffusion coefficient of the particle is a monotonically decreasing or a non-monotonic concave function of obstructed volume fraction. This non-monotonic dependence means that in a range of the parameters, diffusivity of the particle increases with increasing the obstructed volume fraction. Such a counterintuitive effect is shown to result from combination of the effects of the propelling force and anisotropy in the particle shape.
In propelling force-aspect ratio plane, regions corresponding to monotonic and non-monotonic dependence of the diffusion coefficient on the obstacle density are specified both using the simulation results and theoretically. Self-propelling force is found to cause the particle dynamics in the presence of obstacles to consist reentrant diffusive regimes in addition to ballistic and super-diffusive ones.

Let us consider a rod-shaped particle of length $L_{rod}\sigma$, mass $M$ and diameter $d=1\sigma$ (with aspect ratio $L_{rod}$) in three-dimensional space, where $\sigma$ has dimension of length. Configuration of the particle in space can be specified by position of its center, $\vec{r}\equiv (x,y,z)$, and three Euler angles, $\phi$, $\theta$ and $\psi$ \cite{GldStn}. Considering drag and external forces (torques) in addition to the random force (torque) of the thermal noise, the Langevin equations of translational and rotational motions can be written as $M\dot v_i =  - \gamma_i^\textsf{t}v_i + F_i + \eta_i^\textsf{t}(t)$ and $I_i\dot \omega_i =  - \gamma_i^\textsf{r} \omega_i + N_i + \eta_i^\textsf{r}(t)$ where $i = { \bot _1},{ \bot _2},\parallel$ (see Fig. \ref{fig1}) and $"\textsf{t}"$ and $"\textsf{r}"$ stand for translational and rotational, respectively. For a self-propelling particle, it is assumed that $F_i$ contains the self-propelling force as well.
In the above equations, $v_i=\vec{v}.\hat{e}_i$ and $\omega_i=\vec{\omega}.\hat{e}_i$ where $\vec{v}$ and $\vec{\omega}$ are linear and angular velocities of the rod in the lab frame. The noise terms follow $\langle \eta_i^k (t) \rangle  =0$, $\langle \eta_i^k (t)\eta_j^l (t') \rangle  = {\xi_i^k}^2 \delta_{kl}\delta_{ij}\delta (t - t')$ and ${\xi_i^k}^2=2\gamma_i^k k_BT$, where $k=\textsf{t},\textsf{r}$. For motion of a rod-shaped particle in a fluid, it is known that $\gamma _\bot ^\textsf{t}  = 2\gamma _\parallel ^\textsf{t}$ in which $\gamma _ \bot ^\textsf{t}$ ($\gamma _\parallel ^\textsf{t}$) is the friction coefficient the rod experiences when it moves perpendicular (parallel) to its axis \cite{doi1988theory}.
The relation between lab- and body-frame views of any vector $\vec{A}$ reads ${{\vec A}_{body}} = \Re (\phi ,\theta ,\psi ){{\vec A}_{lab}}$, in which $\Re$ is the rotation matrix \cite{GldStn}.
Regarding that $\Re$ has a singularity at $\theta=0$, Cayley-Klein parameters can be used; ${q_1} = \sin (\frac{\theta }{2})\cos (\frac{{\phi  - \psi }}{2})$, ${q_2} = \sin (\frac{\theta }{2})\sin (\frac{{\phi  - \psi }}{2})$, ${q_3} = \cos (\frac{\theta }{2})\sin (\frac{{\phi  + \psi }}{2})$ and ${q_4} = \cos (\frac{\theta }{2})\cos (\frac{{\phi  + \psi }}{2})$. These parameters follow equations $\sum\limits_{i = 1}^4 {q_i^2}  = 1$ and $\sum\limits_{i = 1}^4 {{q_i}{{\dot q}_i}}  = 0$ \cite{GldStn, Rapaport, Singular, Quaternions}.
Direction of the particle in space can be updated as $\vec {q_i}(t + \Delta t) = \vec {q_i}(t) + \vec {\dot {q_i}}(t)\Delta t, i = 1,2,3,4$
in which,  $\vec {\dot q} = Q(\vec q)\vec w$ and $ Q(\vec q)$ is a matrix formed by $q_i$ as:
$$
Q(\vec{q}) = \frac{1}{2}\left[ {\begin{array}{*{20}{c}}
{{q_4}}&{{q_3}}&{ - {q_2}}&{ - {q_1}}\\
{ - {q_3}}&{{q_4}}&{{q_1}}&{ - {q_2}}\\
{{q_2}}&{ - {q_1}}&{{q_4}}&{ - {q_3}}\\
{{q_1}}&{{q_2}}&{{q_3}}&{{q_4}}
\end{array}} \right]
$$

To study obstructed diffusion of the particle, the spherical obstacles were modeled by fixed non-overlapping spheres of diameter $d=1\sigma$,
randomly distributed inside the simulation box. Excluded volume interactions between
the rod and the obstacles were modeled by shifted and truncated sphere-sphere Lennard-Jones potential (see Fig. \ref{fig1}).
The simulation box was a cube of edge length $L=20 \sigma$ and periodic boundary conditions
were applied. Second order Runge-Kutta algorithm was used for integration of the equations of motion \cite{RKII}. The temperature was kept fixed at $T=\frac{1}{k_B}$. As the fluid was modeled implicitly by stochastic and damping terms in the Langevin equation, hydrodynamic interactions were not considered. MD time step was $\tau=0.001\tau_0$, in which
$\tau_0=\sqrt{\frac{M\sigma^2}{k_BT}}$ is the MD time
scale.
The values of the friction coefficients were taken as $\gamma _\parallel ^\textsf{t} = L_{rod}$ (in units of $\sqrt{\frac{Mk_BT}{\sigma^2}}$) , $\gamma _\bot ^\textsf{t} = 2\gamma _\parallel ^\textsf{t}$ and  $\gamma_{\bot}^\textsf{r}=\frac{L_{rod}^2\sigma^2}{6}\gamma _\parallel^\textsf{t}$ \cite{PhysRevE.82.031904}.
For given values of $L_{rod}$, magnitude of the self-propelling force ($F_{sp}$ in units of $\frac{k_BT}{\sigma}$) and the obstructed volume fraction ($\Phi$), the simulations were repeated with 10 different realizations of the obstacles and the averages were calculated. $\Phi$ is defined as the fraction of the simulation box forbidden for a spherical particle of diameter $\sigma$ due to excluded volume interactions.
\begin{figure}[b]
\includegraphics[width=.9 \columnwidth]{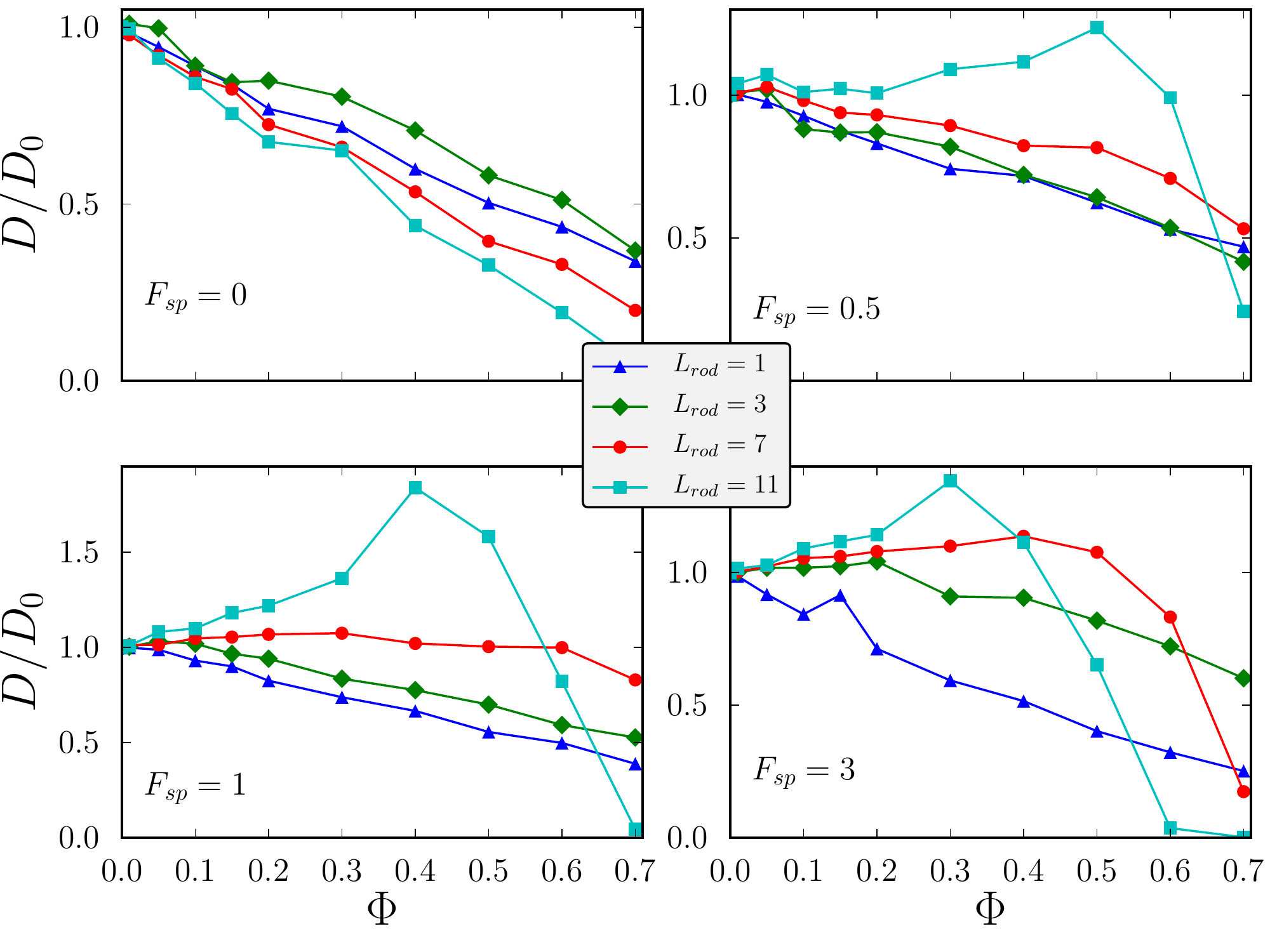}
\caption{(Color online) Reduced diffusion coefficient of particles of aspect ratio $L_{rod}$ versus $\Phi$ for different values of $F_{sp}$. Relative error on the data points is $\simeq 10$ percent. } \label{fig3}
\end{figure}

By simply looking at the trajectories of particles of different $L_{rod}$ at different values of $F_{sp}$ and $\Phi$, an interesting phenomenon can be seen.
With some values of $F_{sp}$ and $L_{rod}$ and in a given duration of time, the SP particle covers considerably longer trajectory in the presence of obstacles relative to the space without obstacles (see Fig. \ref{fig1}, right panel). This is a counterintuitive phenomenon as it is expected the obstruction to decrease the particle diffusivity. Sample trajectories of a sphere of diameter $d=1\sigma$ ($L_{rod}=1$) and a rod of $L_{rod}=11$ obtained from simulations of time length $3\times 10^5 \tau_0$ are shown in Fig. \ref{fig1}. As it can be seen, diffusivity enhancement due to the presence of obstacles doesn't happen for SP spherical particle showing that anisotropy in the particle shape is a key parameter in obstruction-induced enhancement of the diffusivity.
\begin{figure}[b]
\includegraphics[width=.9 \columnwidth]{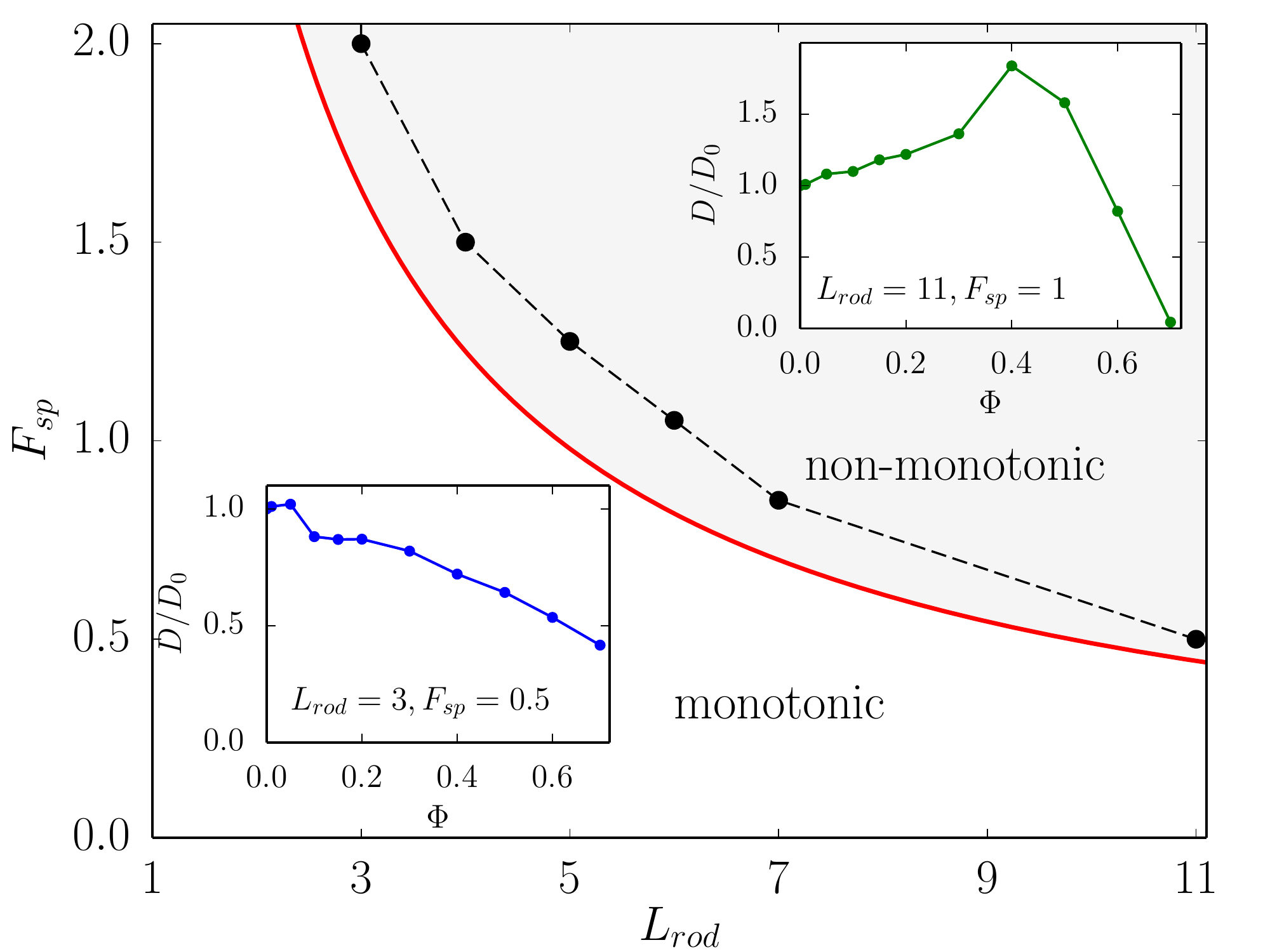}
\caption{(Color online) Regions in $F_{sp}-L_{rod}$ plane in which the diffusion coefficient of the SP rod is a monotonically decreasing or a non-monotonic concave function of obstructed volume fraction. The solid red line shows the theoretical prediction of the boundary between the two regions; $F_{sp}=2\sqrt{6}\frac{k_BT}{L_{rod}}$ and the circle symbols (connected by dashed line as a guide for eyes) are the simulation results for the same boundary.  } \label{fig4}
\end{figure}

As the main measure of the particle diffusivity, we calculate its diffusion coefficient from time dependence of the mean-squared displacement (MSD) and then average over obstacles realizations. In Fig. \ref{fig3}, dependence of the reduced diffusion coefficient, $\frac{D}{D_0}$, on $\Phi$ is shown for different values of $F_{sp}$ and $L_{rod}$ ($D_0$ is the diffusion coefficient in the absence of the obstacles). As it can be seen, for large enough magnitude of the propelling force and the particle aspect ratio, the diffusion coefficient is a non-monotonic concave function of $\Phi$. By addition of obstacles to the environment, the particle diffusivity goes beyond that in the environment empty of obstacles and reaches to a maximum and eventually decreases to zero. For the spherical particle however, $D$ is observed to be a decreasing function of $\Phi$, regardless of the value of $F_{sp}$. In fact, both anisotropy in the particle shape and self-propulsion of enough strength are needed the enhanced diffusivity to occur.

In addition to the translational diffusion coefficient, we calculated orientational relaxation time of rod-shaped particles, $\tau_r$, by looking at their orientational diffusion and from time dependence of $\langle \hat{e}_\parallel (t).\hat{e}_\parallel (0) \rangle$ in the lab frame. It is found that the particle orientation relaxes exponentially as $\exp{(-t/\tau_r)}$ from which $\tau_r$ can be calculated. Also, such orientational relaxation means that $\langle \vec{F}_{sp}(t).\vec{F}_{sp}(0)\rangle={F_{sp}}^2\langle \hat{e}_\parallel (t).\hat{e}_\parallel (0) \rangle={F_{sp}}^2\exp{(-t/\tau_r)}$.

\begin{figure}[b]
\includegraphics[width=.9 \columnwidth]{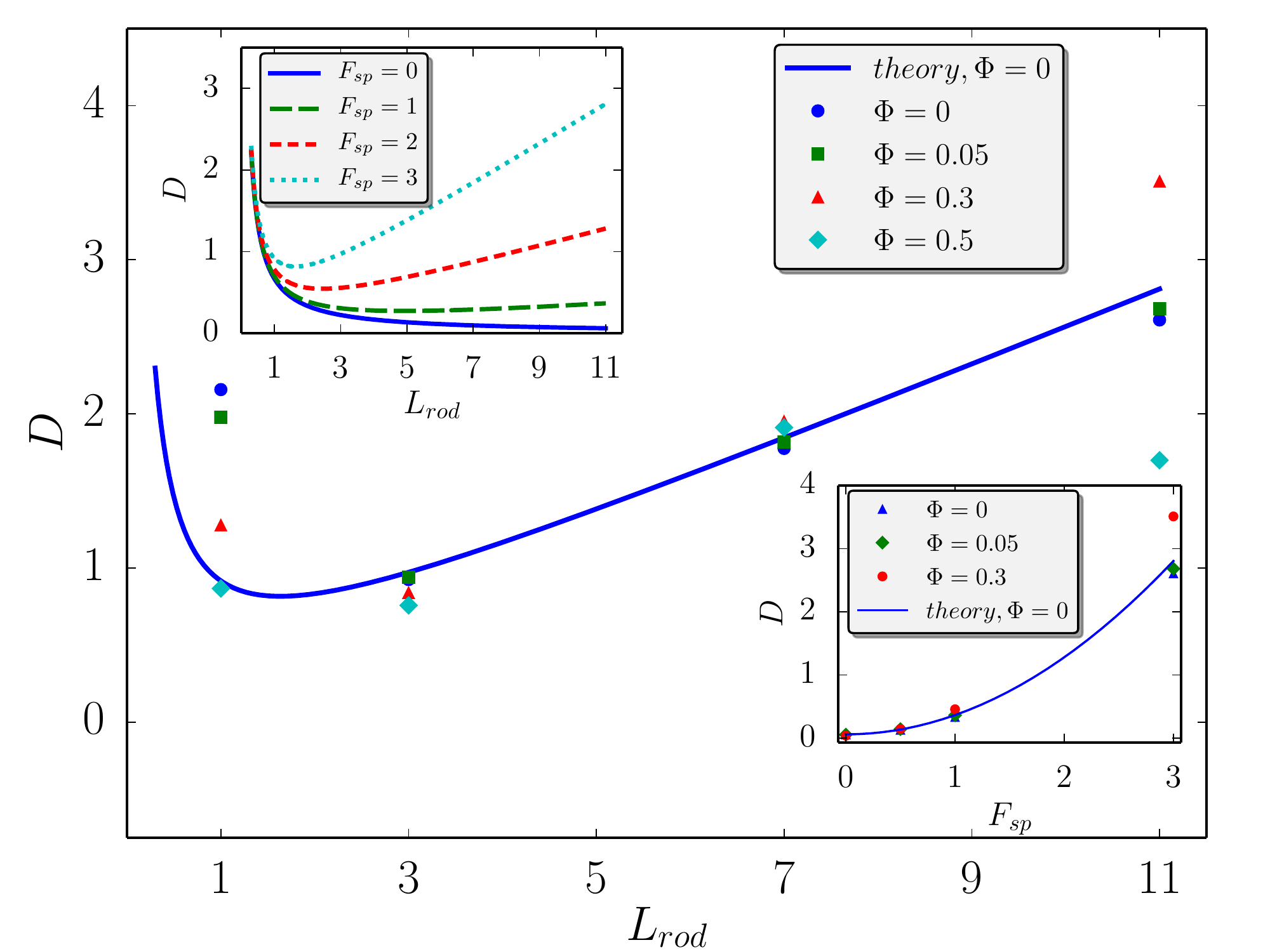}
\caption{(Color online) Diffusion coefficient of a SP particle $(F_{sp}=3)$ versus its aspect ratio. Solid line corresponds to Eq. \ref{D} and the simulation results are shown by symbols. Insets: $D$ versus $L_{rod}$ from Eq. \ref{D}. unlike the passive diffusion for which $D$ is a decreasing function of $L_{rod}$, with nonzero values of $F_{sp}$, $D$ is a convex function of $L_{rod}$ (up). Dependence of $D$ on $F_{sp}$ for a rod of $L_{rod}=11$ (down). } \label{fig5}
\end{figure}

Considering that the values of $L_{rod}$ and $F_{sp}$ determine the reduced diffusion coefficient to be monotonic or non-monotonic function of $\Phi$, a question is how regions of $F_{sp}-L_{rod}$ plane correspond to the two kind of dependencies. The simulation results for answering this question are summarized in Fig. \ref{fig4}. To describe the simulation results shown in Fig. \ref{fig4}, we first calculate the diffusion coefficient of a SP rod in the absence of obstacles, analytically.
To this end, the Langevin equation for the translational degrees of freedom at long enough time scales where the inertial term is washed out, should be solved. In this regime, $v_i(t)=\frac{1}{\gamma_i^\textsf{t}}(\eta_i^\textsf{t}(t)+F_{sp}(t)\delta_{i,\parallel})$ where $i=\parallel, \bot_1, \bot_2$. Therefore, $\Delta x_i(t)=x_i(t)-x_i(0)=\int_0^t v_i(t')dt'$ and $\langle \Delta x_i(t) \Delta x_j(t)\rangle=\frac{1}{\gamma_i^\textsf{t}\gamma_j^\textsf{t}}\int_0^t dt'\int_0^t dt'' [\langle \eta_i^\textsf{t}(t')\eta_j^\textsf{t}(t'')\rangle + \langle F_{sp}(t')F_{sp}(t'')\rangle \delta_{i,\parallel}\delta_{j,\parallel}]$. Considering $\langle F_{sp}(t')F_{sp}(t'')\rangle={F_{sp}}^2\exp(-|t'-t''|/\tau_r)$ and $MSD=\sum_{i=\parallel, \bot_1, \bot_2}\langle\Delta x_i(t)^2\rangle=6Dt$, $D$ can be obtained as $D=\frac{2k_{B}T}{3\gamma_{\parallel}^\textsf{t}}+\frac{\tau_{r}}{3{\gamma_{\parallel}^\textsf{t}}^{2}}{F_{sp}}^2$. From the above equation and that $\tau_r=\frac{1}{2D_r}=\frac{{\gamma_{\parallel}^\textsf{t}}^3}{12k_BT}$,
\begin{equation} \label{D}
 D=\frac{2k_BT}{3\gamma_{\parallel}^\textsf{t}}+\frac{\gamma_{\parallel}^\textsf{t}}{36k_BT}{F_{sp}}^2.
\end{equation}
The first term of $D$ in Eq. \ref{D}, which is a decreasing function of $L_{rod}$, corresponds to the passive diffusion of the particle. Its dependence on $\gamma_{\parallel}^\textsf{t}$ comes from relations $D=\frac{D_{\parallel}+2D_{\bot}}{3}$, and $\gamma_\bot^\textsf{t}=2\gamma_\parallel^\textsf{t}$ \cite{doi1988theory}.
The second term of $D$ in Eq. \ref{D} is the contribution of self-propulsion. Sum of the two terms shows that $D$ is a non-monotonic convex function of $\gamma_\parallel^\textsf{t}=L_{rod}$ with a minimum at $L_{rod}^*=2\sqrt{6}\frac{k_BT}{F_{sp}}$ as shown in Fig. \ref{fig5}.
In this figure, the simulation results for $\Phi=0$ (for comparison with Eq. \ref{D}) and three nonzero values of $\Phi$ are shown. Also, dependence of $D$ on $F_{sp}$ at $\Phi=0$ and two nonzero obstacle densities are shown in Fig. \ref{fig5}. Non-monotonic dependence of active particle diffusivity on its size has also been reported and discussed in Ref. \cite{s-active2}.

The boundary between the two regions of the $F_{sp}-L_{rod}$ plane shown in Fig. \ref{fig4} can be described by considering the non-monotonic dependence of $D$ on $L_{rod}$, shown in Fig. \ref{fig5}. It seems that this boundary corresponds to the relation between $F_{sp}$ and $L_{rod}$ that minimizes $D$; $F_{sp}=2\sqrt{6}\frac{k_BT}{L_{rod}^*}$. For each given value of $F_{sp}$, $D$ has different dependencies on the particle aspect ratio for $L_{rod}<L_{rod}^*$ and $L_{rod}>L_{rod}^*$ (see Figs. \ref{fig4} and \ref{fig5}). For values of aspect ratio, $L_{rod}<L_{rod}^*$, the diffusion coefficient at $\Phi=0$ is a decreasing function of $\gamma_\parallel^\textsf{t}$. In this region, addition of obstacles to the system causes the friction coefficient to increase effectively and hence the diffusion coefficient to be a decreasing function of $\Phi$. For values of aspect ratio, $L_{rod}>L_{rod}^*$, however, the leading term of $D$ is the second term in Eq. \ref{D} which comes from self-propulsion. This term is an increasing function of $\gamma_\parallel^\textsf{t}$ and addition of obstacles has a completely different effect on $D$ relative to the case $L_{rod}<L_{rod}^*$. For $L_{rod}>L_{rod}^*$, addition of obstacles (increasing $\Phi$ from zero), first causes $\tau_r$ and hence $D$ to increase. Addition of more obstacles to the environment eventually causes suppression of the particle motion and decrease of the diffusion coefficient. As it can be seen in Fig. \ref{fig4}, the theoretical prediction of the boundary between the two regions agrees well with the simulation result.

\begin{figure}
\includegraphics[width=.9 \columnwidth]{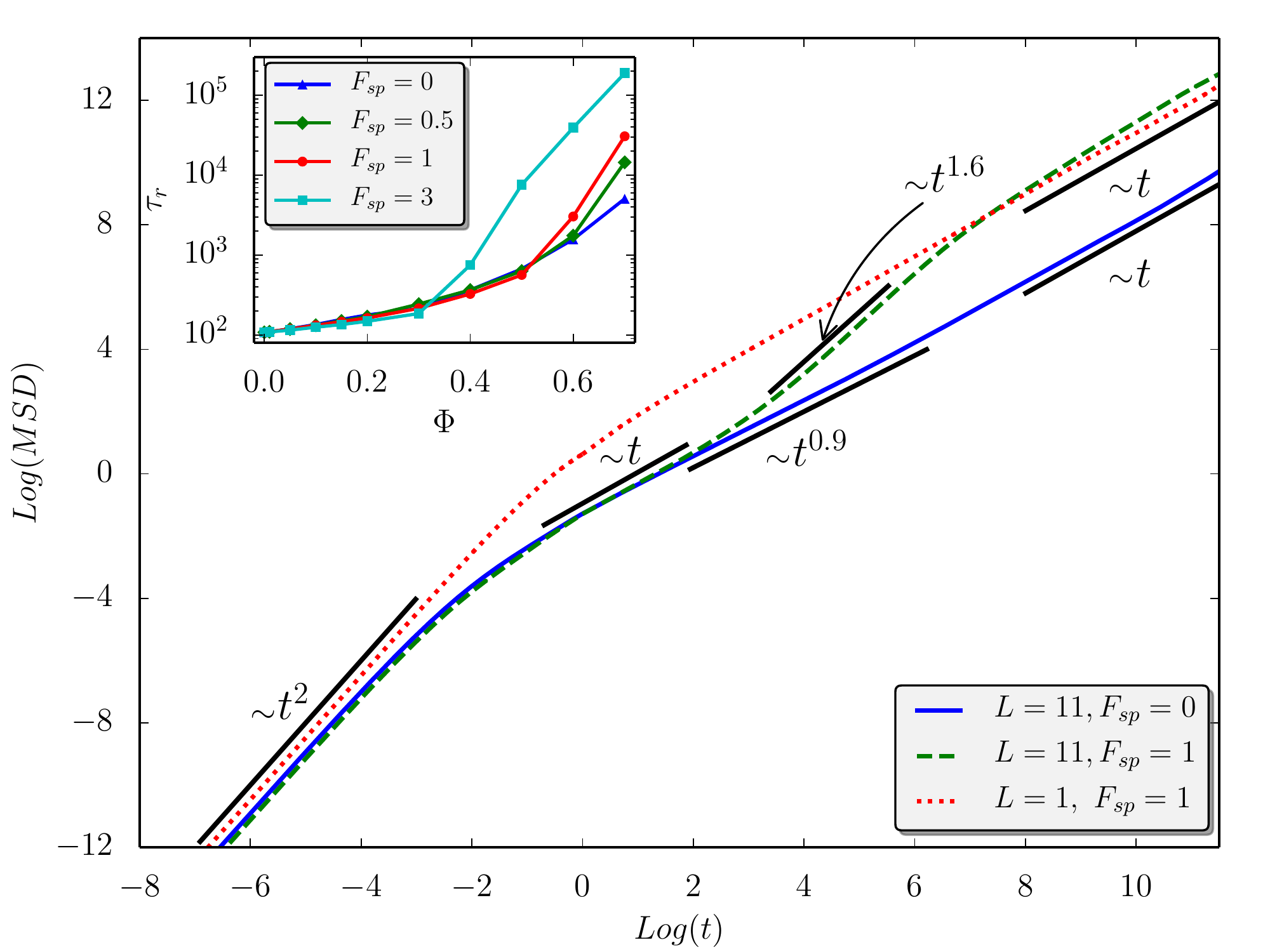}
\caption{(Color online) Log-log plot of MSD versus time for obstructed diffusion of a passive and a SP rod of $L_{rod}=11$ and a SP sphere at $\Phi=0.4$. Inset: Orientational relaxation time versus $\Phi$ for a rod of $L_{rod}=11$. } \label{fig2}
\end{figure}

For a sphere and a rod-shaped particle of aspect ratio $L_{rod}=11$, MSD versus time is shown in Fig. \ref{fig2}. Passive dynamics of the rod consists of ballistic, sub-diffusive and diffusive regimes at short, intermediate and long times, respectively. Self-propelling force causes a supper-diffusive regime to appear in the particle dynamics at intermediate time scales. In addition, reentrant diffusive regimes in the particle dynamics can be seen. Similar behavior has been reported for SP particles in two dimensions \cite{Peruani910}.
The diffusive regime at intermediate time scales is related to the motion of the rod before relaxation of its orientation. Long-time diffusive regime from which we calculate the diffusion coefficient, corresponds to time scales that the rod forgets its orientation and experiences random walk.
In Fig. \ref{fig2}, for a rod of aspect ratio $L_{rod}=11$, $\tau_r$ versus $\Phi$ is shown at four different values of $F_{sp}$.
As it can be seen, $\tau_r$ is a rapidly increasing function of $\Phi$, $F_{sp}$ and $L_{rod}$.

In a stochastic model for nucleosome sliding along DNA under an external force, non-monotonic dependence of the diffusion coefficient on ligand concentration has been reported in Ref. \cite{Farshid}. It has been shown in this work that non-monotonic dependence of the diffusion coefficient on the ligand concentration happens only for forced diffusions and for large enough magnitude of the driving force. In fact, one dimensional nature of the model studied in Ref. \cite{Farshid} provides an inherent spacial anisotropy and makes this model very similar to the motion of a SP rod obstructed by fixed particles. In another model for study of the driven diffusion of particles in the presence of obstacles, non-monotonic dependence of the diffusion coefficient on the driving force has been observed \cite{nonPRE, nonJCP2010, nonJCP2010_2, nonJCP2013}. General similarity between the mentioned models and the problem we studied here originates from the fact that in all of them, two factors namely the driving/self-propelling force and the obstruction compete with each other.

In summary, we have shown that obstruction may enhance the diffusivity of SP particles. Depending on the parameters such as the particle aspect-ratio and the magnitude of the self-propelling force, maximum diffusivity can be achieved by tuning the obstacle density. Regarding that the fabrication of SP particles of various shapes is possible nowadays, the introduced phenomenon could be of interest for drug delivery and many other applications.

\acknowledgements

\end{document}